# Towards Error Correction for Computing in Racetrack Memory

Preston Brazzle, Benjamin F. Morris III, Evan McKinney, Peipei Zhou, *Senior Member, IEEE,*
Jingtong Hu, *Senior Member, IEEE,* Asif Ali Khan, and Alex K. Jones, *Fellow, IEEE*

*Abstract*—Computing-in-memory (CIM) promises to alleviate the Von Neumann bottleneck and accelerate data-intensive applications. Depending on the underlying technology and configuration, CIM enables implementing compute primitives in place, such as multiplication, search operations, and bulk bitwise logic operations. Emerging nonvolatile memory technologies such as spintronic *Racetrack memory* (RTM) promise not only unprecedented density but also significant parallelism through CIM. However, most CIM designs, including those based on RTM, exhibit high fault rates. Existing error correction codes (ECC) are not homomorphic over bitwise operations such as AND and OR, and hence cannot protect against CIM faults. This paper proposes CIRM-ECC, a technique to protect data-intensive RTMs against CIM faults. At the core of CIRM-ECC, we use a recently proposed RTM-based CIM approach and leverage its peripheral circuitry to our implement our novel ECC codes. We show that CIRM-ECC can be applied to single-bit Hamming codes as well as multi-bit BCH codes.

*Index Terms*—Racetrack memory, computing in memory, fault tolerance, error correction codes

## I. INTRODUCTION

Computing-in-memory (CIM) has emerged as a promising approach to mitigate the Von Neumann bottleneck in traditional computing architectures [1]. Most CIM solutions utilize memory devices and their sensing circuits as processing elements. In particular, proposals that leverage DRAM [2], [3] and various non-volatile memories (NVMs) [4], [5] employ triple row activation or specialized access mechanisms to implement logic and arithmetic operations across rows of memory. The accuracy of these computed results often relies on the accuracy of the sensing circuits, as CIM significantly reduces the sensing margins, potentially resulting in high fault rates, such as $10^{-1}$ [6] in a recently proposed DRAM-based CIM accelerator.

Racetrack memory (RTM) is an emerging NVM that offers a promising alternative to other NVMs and traditional memories such as SRAM and DRAM [7]. Like other technologies, RTM facilitates in-place execution of logic and compute operations [5], [8], [9]. CIM in RTM can be performed using different cell attributes, for example, using the resistance state of the access ports [10] or multi-cell access mode called *transverse read* (TR) operation. TR counts the number of '1's across all operands. This enables parallel computation of bitwise operations such as AND (NAND), OR (NOR), and XOR (XNOR). TR accuracy is highly dependent on the precision of the sense amplifier precision due to the narrow sensing margins required to detect the difference between each possible count of '1's [11], [12].

Despite the efficiency of high-density error correction codes (ECC) in protecting memory against individual row faults and chip failures (e.g., Chipkill [13]), codes like Hamming [14] (ECC-1) and BCH [15] (ECC-2 and higher) cannot protect AND and OR operations. These fundamental operations are frequently employed in CIM accelerators [2]–[5] but do not generate homomorphic ECC codes. Consequently, when memory rows containing ECC parity bits are used to execute these bulk bitwise operations, there is no straightforward method to similarly combine the parity bits to ensure protection of the resultant values.

Currently, the leading method of protecting CIM operations is to use $n$-modular redundancy, which requires $n$ copies of the computation to determine the correct result [2], [5]. The value of $n$ depends on theintrinsic and acceptable fault rates. However, this approach divides the potential parallelism of the CIM device by a factor of $n$ (either spatially or temporally), greatly limiting the performance benefits of CIM.

In this paper, we propose CIRM-ECC, or Computing In Racetrack Memory–Error Correction Coding, to protect transverse read-based CIM operations in RTMs. Many ECC methods, including Hamming and BCH codes, are homomorphic over XOR. Thus, two memory rows, data and parity bits for traditional ECC, may be combined with an XOR operation while maintaining data protection. Other logic operations, such as AND and OR, can be protected by concurrently computing XOR. ECC applied to the XOR function can detect faults, correct faults, or, in some cases, determine that the computation was error-free in the primary operation (e.g., AND). This is possible because TR outputs a count of the number of '1's. Faults in a TR will only increase/decrease the '1's count by one. Because bulk bitwise operations are determined by this '1's count and a small amount of additional logic, the output of XOR will switch odd/even parity, which we leverage to protect other logic operations. We demonstrate that CIRM-ECC significantly reduces performance and energy overhead compared to existing $n$-modulo-redundancy-based approaches, while achieving comparable uncorrectable fault rates.

P. Brazzle, E. McKinney, P. Zhou, and J. Hu are with the Department of Electrical and Computer Engineering, University of Pittsburgh, Pittsburgh, PA, 15261 USA e-mail: ({prb50,evm33,peipei.zhou,jthu}@pitt.edu).

B. F. Morris is with the Department of Electrical and Computer Engineering, Duke University, Durham, NC e-mail: ben.morris@duke.edu

A. K. Jones is with the Department of Electrical Engineering and Computer Science, Syracuse University, Syracuse, NY 13244 e-mail: akj@syr.edu

A. A. Khan is with TU Dresden, 01062 Dresden, Germany e-mail: asif_ali.khan@tu-dresden.de

This work is partially supported by NSF under awards CNS-2133267, CNS-2328972, DESC-2324864, and CNS-1822085, and the German Research Council (DFG) through the CO4RTM (450944241) project.

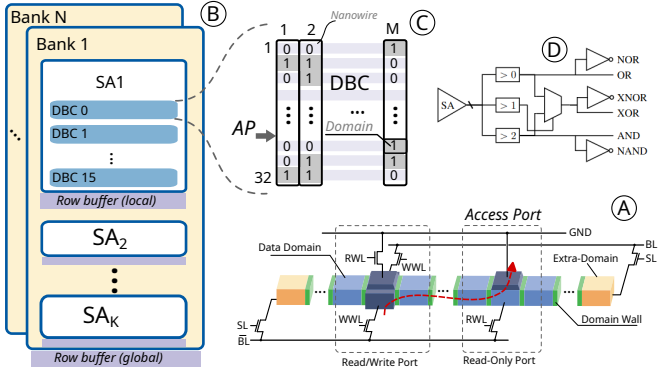

Fig. 1. RTM overview. Ⓐ RTM nanowire showing access ports and associated circuitry; Ⓑ Hierarchical organization; Ⓒ DBC structure, showing $M$ nanowires each storing 32 data bits; Ⓓ Senseamp circuitry for RTM-based CIM; adopted from CORUSCANT [5] for $TRD = 3$.

## II. BACKGROUND AND RELATED WORK

RTM is a spintronic NVM wherein each cell is a magnetic nanowire capable of storing tens of data bits and is equipped with one or more access ports, as shown in Fig. 1A. During read/write operations, the desired data must be *shifted* to the nearest access port by applying a shifting current to an extremity of the nanowire. The access ports typically consist of an access transistor and a magnetic tunnel junction (MTJ) formed with a fixed layer adjacent to one domain of the nanowire serving as the free layer. The alignment of the data with respect to the MTJ determines the bit value of the data, '0' or '1'.

RTMs are hierarchically organized into ranks, banks, and subarrays. Each subarray is separated into independently accessible groups called domain-wall block clusters (DBCs) (Fig. 1B). DBC nanowires are shifted in a lock-step fashion (see Fig. 1C). Data are bit-interleaved across nanowires, allowing access to a data word in parallel. Fig. 1B shows a $512 \times M$ bit subarray with 16 DBCs where each DBC is composed of $M$ 32-bit long RTM nanowires.

**CIM using RTM:** Each nanowire can also function as a polymorphic gate across multiple bits accessed as a count of the '1's (or '0's) within a portion of the nanowire. Two methods, TR [16] or a multi-domain MTJ [11] allow this. The leading CIM method [5], [9] to compute bulk-bitwise logic leverages the TR operation[1]. A TR is conducted between two access points, or an access port and an extremity, for all nanowires within a DBC, as shown by the red arrow in Fig. 1 part C. Note that TR is constrained by the distance between the ports, referred to as the transverse read distance (TRD), with TRD ≤ 7 [5], [11]. Different '1's counts are determined by comparing the sensed voltage/current with fixed thresholds. This can be used to compute bitwise logic operations and multi-operand addition and multiplication for integers [5], and floating point values [9].

TR reports an approximate 3% difference in resistance under process variation [16]. Using the LLG micromagnetic simulator [17] to verify the TR sense margins [16], the fault

---

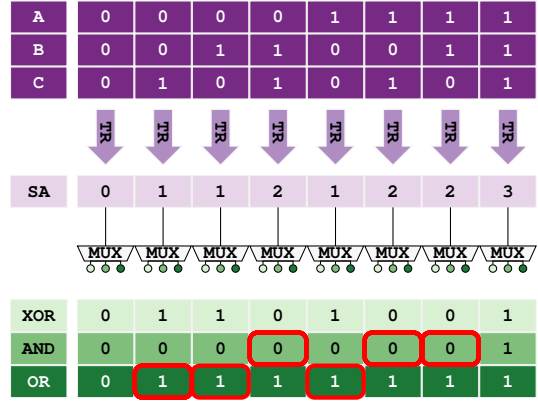

Fig. 2. Example of 3-Input AND and OR Computations in RTM with Ambiguous Fault Locations Outlined

rate for a TR operation is reported to be around $10^{-6}$ for TRD=4. Compared to DRAM-based CIM, this rate is much lower but still requires protection [3]. Although there are ECC schemes for CIM in memristor technology [18]–[21], they are not applicable to RTM. The state-of-the-art RTM-based CIM designs utilize redundancy and voting [5].

## III. CIRM-ECC: PROTECTED RTM-BASED CIM

Fig. 2 illustrates an example of an AND operation using three operands $(A, B, C)$. At each bit position, the TR output at the corresponding senseamp can be 0–3 '1's, depending on the operands' bits. The output of the AND operation is '1' when the TR = 3, as all operands are '1' (rightmost bit position). Similarly, OR is '1' for TR ≥ 1 (all but the leftmost position).

Next, we show how faults arise in AND and OR operations and are detected and corrected using the XOR operation.

### A. XOR computation and homomorphism over ECC

Fig. 1D shows the sensing and logic circuitry for TRD = 3. While AND and OR detect their result from the sensing extremities, XOR is a function of all sense results. Since many ECC schemes, including Hamming, Huffman, Reed-Solomon, and BCH codes, are homomorphic over XOR, the XOR of any two ECC codewords is a valid codeword for the XOR of the two corresponding data words. Therefore, a fault in the XOR operation can be detected and corrected with traditional ECC hardware. In this work, we leverage this property to protect other logic operations.

In addition to its compatibility with existing ECC codes, XOR is useful because of its ability to always detect single-level sensing faults. A sensing fault reports a different number of '1's than stored in the nanowire segment, and is dominated by counts of ±1. These single-level faults will flip the result of XOR. Larger distance faults (two-level and up) are exponentially less likely than single-level faults. Thus, correcting single-level faults is the focus of CIRM-ECC.

### B. Fault handling in non-XOR logic operations

In many cases, faults detected by XOR will also show up as errors in other CIM operations. However, there are instances where an faults detectable through XOR do not create errors in other CIM operations. Depending on the sense level and whether a fault is detected through `XOR` it may be possible to infer that the flip will or will not affect the result of `AND` and/or `OR` operations. In other cases, it cannot be known if the fault led to a error. Thus, sensing faults detected by `XOR` can lead to deterministic errors, deterministic non-errors, or be ambiguous faults. Deterministic cases can be directly corrected or left alone, while ambiguous cases must be recomputed.

---

[1] RTM CIM [5], [9] conceptually is compatible with either TR [16] or a multi-domain MTJ [11], we presume TR for the remainder of this paper.

Considering the CIM circuit in Fig. 1 part D and our three-operand example in Fig. 2, if the senseamp detects a fault and reads zero '1's, the correct count should be one '1'. This is a deterministic error for `OR`, such that the result can be corrected from '0' to '1'. It is a deterministic non-error for `AND` such that the '0' result remains unchanged. Thus, this fault can be directly corrected (if needed) so that we return the correct value. However, if the senseamp falsely produces a single '1' as output, the `XOR` output will report a fault, but it is unclear whether the correct count was zero '1's or two '1's. In this case, the `AND` result remains a deterministic non-error returning a '0' result. However, for the `OR` operation it would not be possible to infer the correct result (an ambiguous fault) as zero '1's is a possibility and would return a '0' instead of '1' requiring the CIM operation to be conducted again.

Fig. 2 shows the type of fault for each possible senseamp reading with three operands. For the `AND` operation, the `XOR` fault is ambiguous when two '1's are reported, while all other cases are not errors for `AND`, as the output must be '0' regardless. Similarly, for the `OR` operation, the fault is ambiguous when reporting a single '1'. Generally, for $n$ operands, a fault is ambiguous for `AND` when the senseamp reports $n-1$ '1's and for `OR` when it reports a single '1'. These cases are highlighted with red circles in Fig. 2. Deterministic errors occur when a fault is detected for `AND` when sensing $n$ '1's and for `OR` when sensing zero '1's. Single-level faults will not affect `AND` when sensing fewer than $n-1$ '1's and `OR` when sensing more than a single '1'. Ambiguous faults must be recomputed.

### C. Fault rates calculation

To show the effectiveness of CIRM-ECC, we calculate the fault rates for `AND` and `OR` operations based on the rate of single-level sensing faults. For each operation, there are two sensing fault possibilities: falsely reading too many '1's, and falsely reading too few '1's. However, for `AND` and `OR` operations, there are only limited cases where a fault translates into an error. For `AND` if an input of $n$ '1's is read as too few '1's and if an input of $n-1$ '1's is read as too many '1's. `OR` is the same except with '0's. If the data follow a uniform distribution (each input is equally likely) then for each fault there is a $\frac{1}{n}$ probability that a fault will appear as an error for `AND` or `OR` and, by extension, `NAND` or `NOR`. Thus, the fault rate $P$ for different operations can be represented by Eq. 1 where $P(\text{XOR})$ is the probability of a single-level sensing fault.

$$P(\text{AND}) = P(\text{OR}) = \frac{P(\text{XOR})}{n} \tag{1}$$

These bit-level fault rates can easily be extended to word- and row-level fault rates by modeling the fault of each bit as an independent event. However, for detection and correction it is necessary to use $P(\text{XOR})$ as all single-level faults appear as flips regardless of whether they are a fault in `AND` or `OR`. Thus, the uncorrectable fault rate can be approximated by a binomial distribution of $P(\text{XOR})$ and the detection/correction capability of the ECC scheme. Because not every sensing fault results in an error, the binomial distribution formula represents a worst case uncorrectable error rate for CIRM-ECC based on the ECC scheme applied.

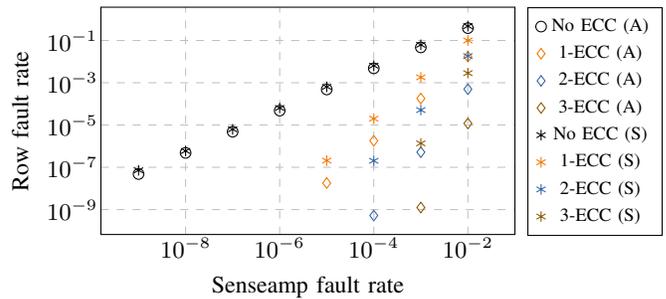

Fig. 3. Comparison of analytical (A) error rates and RTSim (S) error rates for AND-OR operations.

## IV. Evaluation

This section provides an evaluation of CIRM-ECC for different initial fault rates and a comparison with existing non-protected and modular-redundancy-based protection schemes both with stochastic inputs and representative workloads.

### A. Experimental setup

We extended the RTSim [22] simulator to support TR-based CIM operations and to simulate CIM faults[2]. Specifically, for each TR operation, the simulator probabilistically injects faults into each word based on a fault rate parameter. If the number of faults in any word exceeds the ECC correction capacity, it is recorded as an uncorrectable fault, and the instruction is not reissued. If the fault count is within the ECC capacity and are non-ambiguous, they are corrected by the ECC module. Otherwise, the instruction is reissued to account for re-reading the senseamps, and the performance and energy metrics are updated accordingly. Consequently, simulations with higher ECC levels result in more transverse reads, increasing energy consumption while reducing the number of uncorrectable faults. First we evaluate the performance of CIRM-ECC on a series of `AND`/`OR` operations with stochastic inputs. We then evaluate uncorrectable error and performance using three benchmarks: in-memory counters, AES encryption, and matrix-matrix multiplication. All benchmarks are implemented with TR-based `AND`/`OR` operations.

### B. Analytical vs simulated fault rates

The fault rate calculations for 512-bit operands, divided into eight 64-bit words, under various fault rates are illustrated in Fig. 3. The results are generated with a synthetic trace consisting 10M `AND`/`OR` CIM operations. The trace is run

---

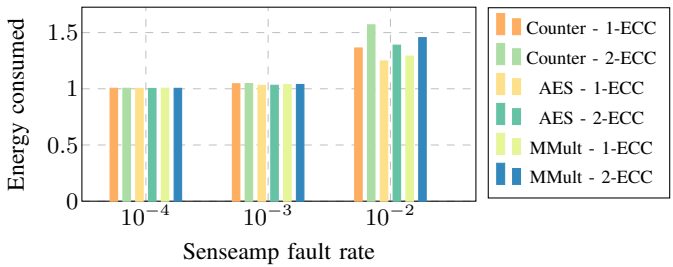

Fig. 4. CIRM-ECC energy consumption normalized to no ECC

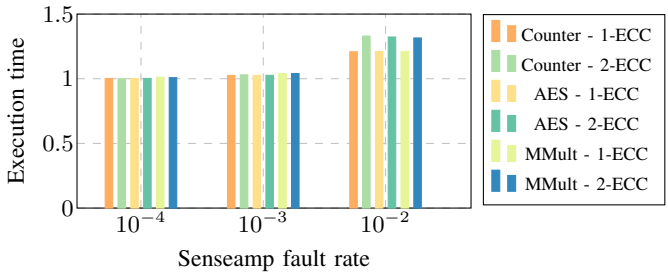

Fig. 5. CIRM-ECC execution time normalized to no ECC

multiple times to make sure faults are injected even at smaller fault rates, e.g., $10^{-9}$.

The No-ECC configuration reflects the probability of a single fault. The 1-ECC case assumes that one fault can be protected, indicating the probability of having more than one fault. The same applies to 2-ECC and 3-ECC. In the 3-MR scenario, a fault in one copy is protected if the remaining two copies are correct, reflecting the probability of having two or more faults. The same applies to 5-MR and 7-MR. Using a 1-ECC code for CIRM-ECC provides comparable protection to 3-MR, such that the 64-72 Hamming code reducing spatial overhead from 200% to just 12.5%. An ECC code that corrects up to three errors is analogous to 5-MR. This demonstrates that CIRM-ECC significantly reduces spatial overhead compared to existing methods.

In the simulation results, the more robust error correction schemes ("2-ECC" and "3-ECC") demonstrate a greater proportional deviation from the analytical results. However, the largest absolute difference is in the "No ECC" series and still only around 9%.

### C. Impact on execution time and energy consumption

Fig. 4 shows the energy overhead associated with different ECC and redundancy schemes, highlighting how CIRM-ECC performs relative to traditional methods. As expected, for higher fault rates more ambiguous faults were detected, requiring reissue of the CIM commands and leading to an average overhead of 39% (>50% for the 2-ECC in the Counter benchmark). In contrast, for the fault rate of $10^{-4}$, only 0.2% of the CIM instructions produced faults, leading to a nominal overhead of 0.4%. For lower fault rates as in prior work [5] the correction overhead would be nominal. The performance overhead presented in Fig. 5 shows a similar overall trend, increasing the execution time by over 30% for 2-ECC at a sensing fault rate $10^{-2}$ and falling off as the fault rate decreases. Note that these results also include the latency and energy consumption of the basic `RD/WR` operations.

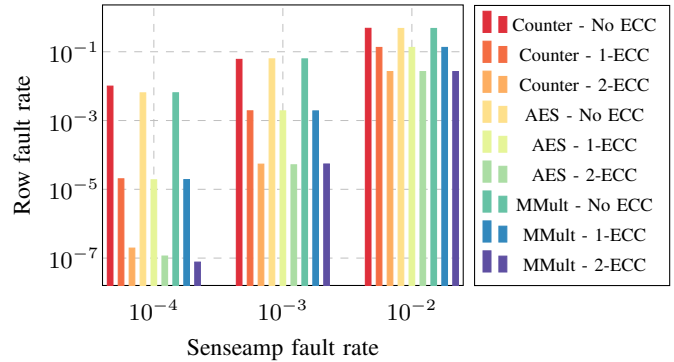

Fig. 6. CIRM-ECC uncorrectable senseamp and row fault rates

Finally, Fig. 6 demonstrates the uncorrectable bit error rate (UBER) for different levels of ECC. Even at the same sensing fault rate the row fault rate does change appreciably across different benchmarks due to different type of CIM operations and different data distributions, leading to different numbers of CIM faults.

## V. Conclusion

Existing error correction codes are not homomorphic over bitwise operations such as `AND` and `OR` common in CIM implementations. This makes traditional high-density codes unsuitable for protecting CIM faults. This paper introduces CIRM-ECC, designed to protect CIM faults in racetrack memory. CIRM-ECC is based on a recently proposed CIM approach that utilizes multi-domain access to execute bulk bitwise operations [5], [9]. This method enables simultaneous computation of CIM operations and parity checking, allowing CIM computation to be performed in tandem with fault detection and correction. CIRM-ECC can be applied to both single-bit Hamming codes and multi-bit BCH codes, offering adaptable fault protection based on the rate of single-level sensing faults.